\documentclass[twocolumn,english,aps,twocolum,superscriptaddress,amsmath,amssymb,floatfix]{revtex4-1}
\usepackage[colorlinks=true,citecolor=blue,linkcolor=magenta]{hyperref}

\usepackage[markup=blue, authormarkupposition=left]{changes} 

\usepackage{soul}
\usepackage[utf8]{inputenc}
\usepackage[english]{babel}
\usepackage{amsmath,amsfonts,amssymb}
\usepackage[T1]{fontenc}
\usepackage{url}

\usepackage{amsmath}
\usepackage{amsfonts}
\usepackage{amssymb}

\usepackage{epstopdf}
\usepackage{graphicx}

\begin{document}

\title{Chip-Based Laser with 1 Hertz Integrated Linewidth}

\author{Joel~Guo}
\thanks{These authors contributed equally to this work.}
\affiliation{Department of Electrical and Computer Engineering, University of California, Santa Barbara, Santa Barbara, California 93106, USA}

\author{Charles~A.~McLemore}
\thanks{These authors contributed equally to this work.}
\affiliation{National Institute of Standards and Technology, 325 Broadway, Boulder, CO 80305, USA}
\affiliation{Department of Physics, University of Colorado Boulder, 440 UCB Boulder, CO 80309, USA}

\author{Chao~Xiang}
\affiliation{Department of Electrical and Computer Engineering, University of California, Santa Barbara, Santa Barbara, California 93106, USA}

\author{Dahyeon~Lee}
\affiliation{National Institute of Standards and Technology, 325 Broadway, Boulder, CO 80305, USA}
\affiliation{Department of Physics, University of Colorado Boulder, 440 UCB Boulder, CO 80309, USA}

\author{Lue~Wu}
\affiliation{T. J. Watson Laboratory of Applied Physics, California Institute of Technology, Pasadena, CA 91125, USA}

\author{Warren~Jin}
\affiliation{Department of Electrical and Computer Engineering, University of California, Santa Barbara, Santa Barbara, California 93106, USA}

\author{Megan~Kelleher}
\affiliation{National Institute of Standards and Technology, 325 Broadway, Boulder, CO 80305, USA}
\affiliation{Department of Physics, University of Colorado Boulder, 440 UCB Boulder, CO 80309, USA}

\author{Naijun~Jin}
\affiliation{Department of Applied Physics, Yale University, New Haven, CT 06520, USA}

\author{David~Mason}
\affiliation{Department of Applied Physics, Yale University, New Haven, CT 06520, USA}

\author{Lin~Chang}
\affiliation{Department of Electrical and Computer Engineering, University of California, Santa Barbara, Santa Barbara, California 93106, USA}

\author{Avi~Feshali}
\affiliation{Anello Photonics, Santa Clara, CA, USA}

\author{Mario~Paniccia}
\affiliation{Anello Photonics, Santa Clara, CA, USA}

\author{Peter~T.~Rakich}
\affiliation{Department of Applied Physics, Yale University, New Haven, CT 06520, USA}

\author{Kerry~J.~Vahala}
\affiliation{T. J. Watson Laboratory of Applied Physics, California Institute of Technology, Pasadena, CA 91125, USA}

\author{Scott~A.~Diddams} 
\affiliation{National Institute of Standards and Technology, 325 Broadway, Boulder, CO 80305, USA}
\affiliation{Department of Physics, University of Colorado Boulder, 440 UCB Boulder, CO 80309, USA}
\affiliation{Department of Electrical, Computer and Energy Engineering, University of Colorado Boulder, 425 UCB, Boulder, CO 80309, USA}

\author{Franklyn~Quinlan}
\email[]{franklyn.quinlan@nist.gov}
\affiliation{National Institute of Standards and Technology, 325 Broadway, Boulder, CO 80305, USA}
\affiliation{Department of Physics, University of Colorado Boulder, 440 UCB Boulder, CO 80309, USA}

\author{John~E.~Bowers}
\email[]{jbowers@ucsb.edu}
\affiliation{Department of Electrical and Computer Engineering, University of California, Santa Barbara, Santa Barbara, California 93106, USA}

\begin{abstract}
Lasers with hertz-level linewidths on timescales up to seconds are critical for precision metrology, timekeeping, and manipulation of quantum systems. Such frequency stability typically relies on bulk-optic lasers and reference cavities, where increased size is leveraged to improve noise performance, but with the trade-off of cost, hand assembly, and limited application environments. On the other hand, planar waveguide lasers and cavities exploit the benefits of CMOS scalability, but are fundamentally limited from achieving hertz-level linewidths at longer times by stochastic noise and thermal sensitivity inherent to the waveguide medium. These physical limits have inhibited the development of compact laser systems with frequency noise required for portable optical clocks that have performance well beyond conventional microwave counterparts. In this work, we break this paradigm to demonstrate a compact, high-coherence laser system at 1548~nm with a 1~s integrated linewidth of 1.1~Hz and fractional frequency instability less than 10$^{-14}$ from 1~ms to 1~s. The frequency noise at 1~Hz offset is suppressed by 11~orders of magnitude from that of the free-running diode laser down to the cavity thermal noise limit near 1~Hz$^2$/Hz, decreasing to 10$^{-3}$~Hz$^2$/Hz at 4~kHz offset. This low noise performance leverages wafer-scale integrated lasers together with an 8~mL vacuum-gap cavity that employs micro-fabricated mirrors with sub-angstrom roughness to yield an optical $Q$ of 11.8~billion. Significantly, all the critical components are lithographically defined on planar substrates and hold the potential for parallel high-volume manufacturing, and the total laser system optics can be housed in a 150~mL custom compact module. Consequently, this work provides an important advance towards compact lasers with hertz-level linewidths for applications such as portable optical clocks, low-noise RF photonic oscillators, and related communication and navigation systems.
\end{abstract}

\maketitle

\section*{Introduction}
Narrow linewidth, frequency stable lasers are essential in precision photonic microwave oscillators and atomic systems including atomic clocks, gyroscopes, and sensors. Particularly, for atomic clocks, they are necessary for cooling, trapping, and probing an atomic species~\cite{ludlow2015optical, boyd2007high}. In this arena, the state of the art has demonstrated laser linewidth below 10~mHz and instability as low as $4\times10^{-17}$~\cite{matei20171}. However, such remarkable performance requires systems with significant size, complexity, and even cryogenic temperatures, all of which restrict their application to laboratory settings. On the other hand, there is significant interest in, and need for, field deployment of atom-based systems~\cite{moody2021roadmap, mehta2020integrated, niffenegger2020integrated}. 
Similarly, state-of-the-art low-noise microwave oscillators based on cryogenic sapphire resonators take advantage of their higher quality factor ($Q$) at ultra-low temperatures~\cite{hartnett2012ultra}, but at the cost of restricting the range of operational environments. Optical frequency division (OFD) takes advantage of higher $Q$s of optical resonators at room temperature and converts stability in the optical domain down to the RF domain via photodetection of a frequency comb~\cite{fortier2011generation}. By stabilizing the comb to an optical reference cavity, the stability of the optical reference is transferred to the microwave signal. However, the best OFD systems also rely on bulk-optic lasers and cavities~\cite{nakamura2020coherent}, impeding the advancement of field applications requiring extraordinarily low microwave phase noise. 

\begin{figure*}[!htb]
\centering
\includegraphics[width=\linewidth]{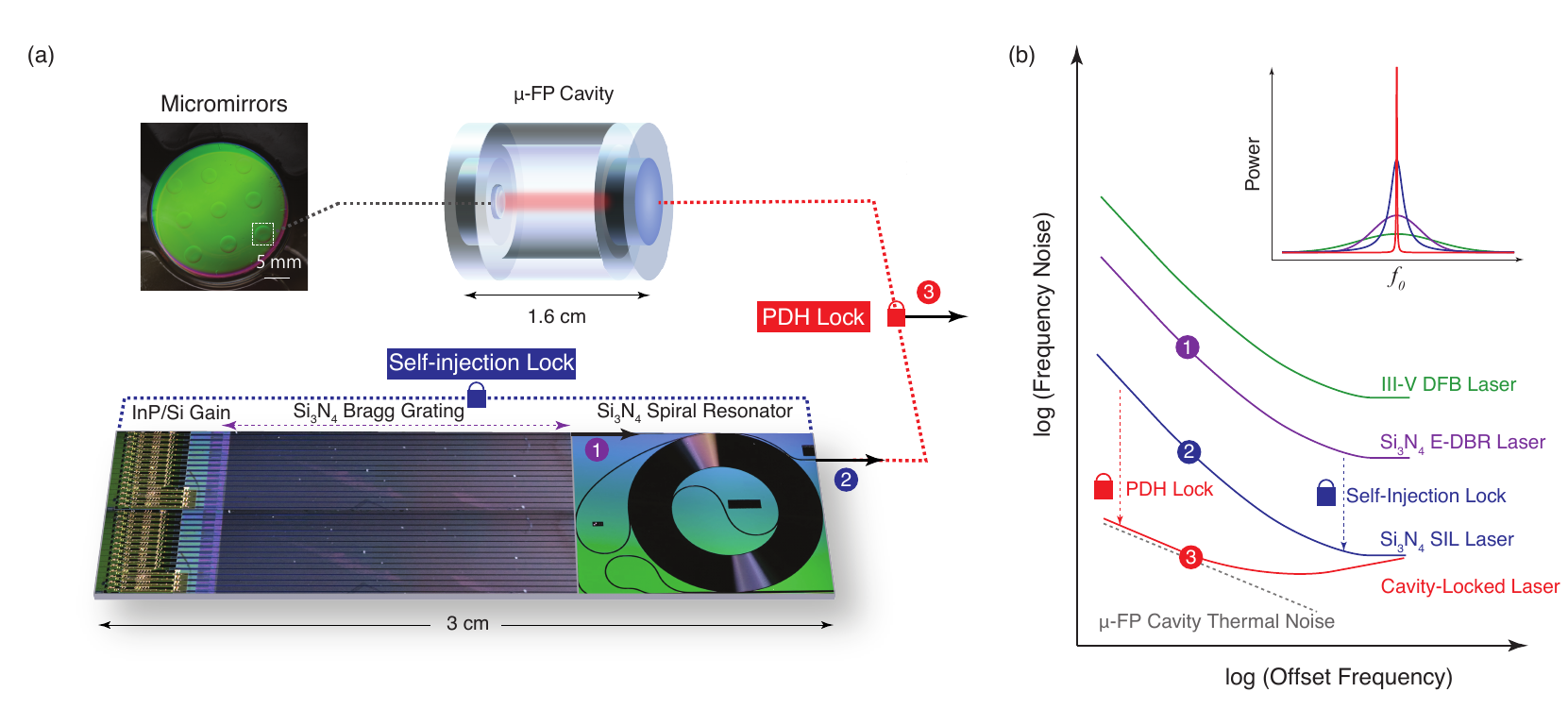}
\caption{Critical components and frequency noise reduction concept of our cavity-stabilized laser. \textbf{(a)} Images of the chip-scale components used in this work, including the laser, spiral resonator, and cavity micromirrors. \textbf{(b)} Concept frequency noise and linewidth reduction from each stage of feedback. The laser source consists of an electrically driven, heterogeneous III-V/Si/Si$_3$N$_4$ laser~\textbf{(1)}, featuring a narrowband integrated Si$_3$N$_4$ extended distributed Bragg reflector (E-DBR), which significantly reduces its white frequency noise floor from that of a monolithic III-V DFB~\cite{xiang2021high}. Self-injection locking to a high-$Q$ Si$_3$N$_4$ spiral resonator further suppresses the frequency noise~\textbf{(2)}, ultimately limited by thermo-refractive noise (TRN)~\cite{jin2021hertz,Li:21}. Subsequent PDH locking to a high-finesse, micro-fabricated vacuum cavity~\cite{mclemore2022thermal, jin2022scalable, hendrie2021versatile} overcomes these limits, drastically reducing the noise down to the cavity thermal noise floor at offsets within the servo bandwidth~\textbf{(3)}. In contrast to bulk-optic lasers and cavities, our laser's critical components are lithographically defined, showing great promise for wafer-level scalability.}
\label{fig:architecture}
\end{figure*}

As the need for ultra-stable optical and microwave sources and atomic systems grows, integrated lasers and photonic circuits provide a compelling path towards system-level integration. This vision combines active (lasers, modulators, detectors), passive (filters, off-chip coupling), and nonlinear elements (frequency combs, frequency converters) while maintaining small overall volume~\cite{komljenovic2016heterogeneous, newman2019architecture, papp2014microresonator, xue2015mode, wang2021self, lu2020chip, domeneguetti2021parametric, chang2019strong, luo2018highly, ropp2021multi, mcgehee2021magneto, spektor2021inverse}.
In addition, heterogeneous silicon photonics offers a path towards realizing ultra-stable, high-precision laser performance in a compact and mobile platform and has demonstrated tremendous scalability with 300~mm wafer-scale fabrication of Tb/s photonic transceivers for data center applications~\cite{jones2019heterogeneously, margalit2021perspective, Xiang2021JSTQE}. Silicon nitride (Si$_3$N$_4$) photonics adds even more functionality, taking advantage of CMOS compatibility, wide bandgap, and low-loss integrated waveguides~\cite{bauters2011ultra, blumenthal2020photonic, xiang2020effects}. Si$_3$N$_4$-based lasers have especially leveraged low-loss~\cite{xiang2021laser, xiang2021high, jin2021hertz, liu2021photonic, puckett2021422} and have demonstrated coherence on par with commercial fiber lasers~\cite{Li:21}. However, they are ultimately limited by thermo-refractive noise (TRN), which has kept the fractional frequency instability of planar waveguide and solid dielectric resonators above the $10^{-13}$ level typical of quartz oscillators~\cite{lee2013spiral}. 

In the experiments reported here, we demonstrate a laser utilizing planar micro-fabricated critical components and exhibiting an integrated linewidth at the 1 Hz level---a value compatible with the performance requirements of compact optical clocks and low-noise RF photonic oscillators. The corresponding frequency instability of the laser is below $1\times10^{-14}$ for averaging times between 1~ms and 1~s, which is significantly below that of the best quartz oscillators~\cite{salzenstein2010significant}.
We overcome TRN limits inherent in planar and other solid dielectric reference cavities by joining a micro-fabricated laser and vacuum-gap Fabry-Perot reference cavity. We lock our integrated self-injection locked (SIL) laser with the Pound-Drever-Hall technique to an 8~mL vacuum-gap cavity~\cite{mclemore2022thermal} formed from lithographically fabricated micromirrors with large, user-defined radius of curvature, pristine surface quality, and high finesse~\cite{hendrie2021versatile, jin2022scalable}. In contrast to planar waveguide and solid dielectric resonators, the optical mode in a vacuum-gap cavity only interacts with matter at the coated dielectric mirror surfaces; by limiting this interaction to the mirrors, we reduce the influence of stochastic fluctuations inherent to all matter at finite temperature and achieve an extremely low thermal noise floor. Similarly compact reference cavities have demonstrated excellent frequency stability but lack a path to scalability and integration~\cite{zhang2020ultranarrow, stern2020ultra, davila2017compact, didier2018ultracompact, leibrandt2011spherical, webster2011force} On the other hand, integrated planar waveguide devices have demonstrated a clear path to efficient scalability but suffer from higher frequency noise~\cite{liu202136, liu2021photonic, puckett2021422}. By utilizing lithographically fabricated micromirrors in our cavity, we capitalize on the noise performance of vacuum-gap systems while introducing parallel manufacturability previously reserved for planar waveguide systems. The performance of our cavity-stabilized laser is therefore unprecedented in chip-based devices, yielding a 1.1~Hz linewidth at 1~s, frequency noise following the cavity noise floor down to 10$^{-3}$~Hz$^2$/Hz at kHz offsets, and Allan deviation better than 10$^{-14}$ averaged out to 1~s.
With such scalable components, we bridge the gap between silicon photonics and the laser performance required to provide new opportunities for applications in high-precision GPS-free positioning, navigation, and timing (PNT), next-generation radar, and commercial 5G communications.

Figure \ref{fig:architecture}a shows images of the critical components used for the experiment. The pump source is a fully integrated and electrically driven heterogeneous III-V/Si/Si$_3$N$_4$ laser, fabricated via wafer bonding on a 100~mm wafer~\cite{xiang2020narrow,xiang2021high}. With a 20~mm long Si$_3$N$_4$ extended distributed Bragg reflector (E-DBR) fully integrated into the laser cavity, the laser exhibits an instantaneous linewidth of 400~Hz with an on-chip output power over 10~mW~\cite{xiang2021high}. Due to the narrow-band feedback of the E-DBR, the instantaneous linewidth corresponding to the white frequency noise floor is significantly reduced (\ref{fig:architecture}b-1) compared to that of a solitary gain section laser typical of monolithic III-V DFB lasers~\cite{tran2019tutorial, tran2019ring}. 

For further noise suppression, the E-DBR laser was edge-coupled and self-injection locked to a high-$Q$ Si$_3$N$_4$ spiral resonator on a separate chip~\cite{Li:21,xiang2021high}. In this scheme, resonant backward Rayleigh scattering is fed back to the laser, which drastically suppresses the frequency noise at offset frequencies within the resonance linewidth~(\ref{fig:architecture}b-2)~\cite{kondratiev2017self, savchenkov2018stiffness, liang2015ultralow}. The noise reduction is proportional to $Q^2$ up to the TRN limit, which depends on the modal volume~\cite{lee2013spiral}. With a measured loaded $Q$ of 126~million and free spectral range (FSR) of 135~MHz, the resonator used for this experiment provides frequency noise superior to that of a commercial fiber laser~\cite{Li:21}. Subsequently, our experiment also demonstrates the viability and advantage of using integrated lasers over fiber lasers in terms of both size, integration, and noise performance. These resonators were fabricated on a 200~mm substrate in a CMOS foundry and feature the same cross-sectional waveguide geometry as the Si$_3$N$_4$ waveguide in the laser, yielding high modal overlap~\cite{jin2021hertz, xiang2021high}. 

The final component to improve noise suppression is a vacuum-gap micro-fabricated Fabry-Perot ($\mu$-FP) cavity constructed with a lithographically defined micromirror~\cite{hendrie2021versatile, jin2022scalable, mclemore2022thermal}. Micro-fabrication allowed curved mirrors to be etched on a fused silica substrate that was then optical-contact bonded to a 10~mm long, 25.4~mm diameter ultra-low expansion (ULE) glass spacer; a fused silica flat mirror was bonded on the opposite side of the ULE spacer. Both mirrors are coated with a highly reflective (>~99.999\%) dielectric stack. The resultant cavity features a finesse of 920~000 ($Q$ of 11.8~billion) and a linewidth of 16~kHz. High finesse is essential for stabilizing light at the thermal noise floor of the cavity, as finesse directly constrains the maximum achievable discriminator slope used in PDH locking~\cite{black2001introduction}. The overall cavity volume is 8~mL; while a small cavity volume is essential for integrated systems, it has the additional benefit of reducing the cavity's sensitivity to external vibrations~\cite{Chen2006vibration}. With careful cavity design and isolation (details in Materials and Methods), we simulate a frequency noise floor of $0.72/f$~Hz$^2$/Hz (phase noise floor of $-4.4/f^3$~dBc/Hz) and observe a cavity drift of a few Hz/s or better over multiple hours.

\begin{figure}[!htb]
\centering
\includegraphics[width=\linewidth]{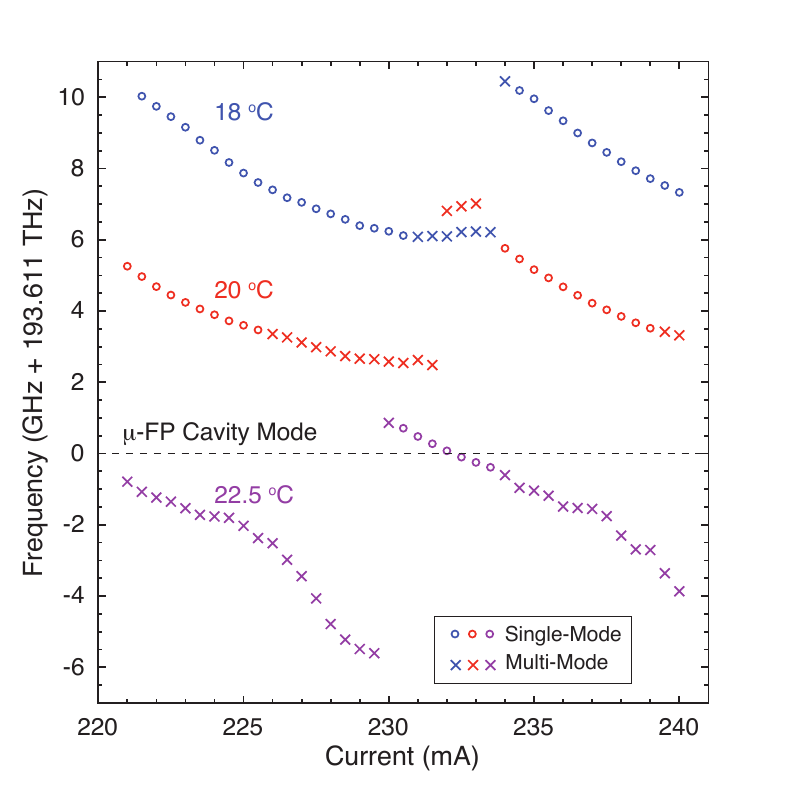}
\caption{E-DBR laser frequency tuning map. The laser current and temperature are tuned across single- and multi-mode regimes. The laser is tuned to $\sim$232~mA and 22.5~$^\circ C$, so that it operates in a single mode at a $\mu$-FP cavity resonance. Laser frequency tuning over half the 15 GHz $\mu$-FP FSR ensures that a $\mu$-FP resonance can always be reached, given arbitrary frequency alignment of the laser and $\mu$-FP cavity.}
\label{fig:tuning}
\end{figure}

\begin{figure*}[!htb]
\centering
\includegraphics[width=\linewidth]{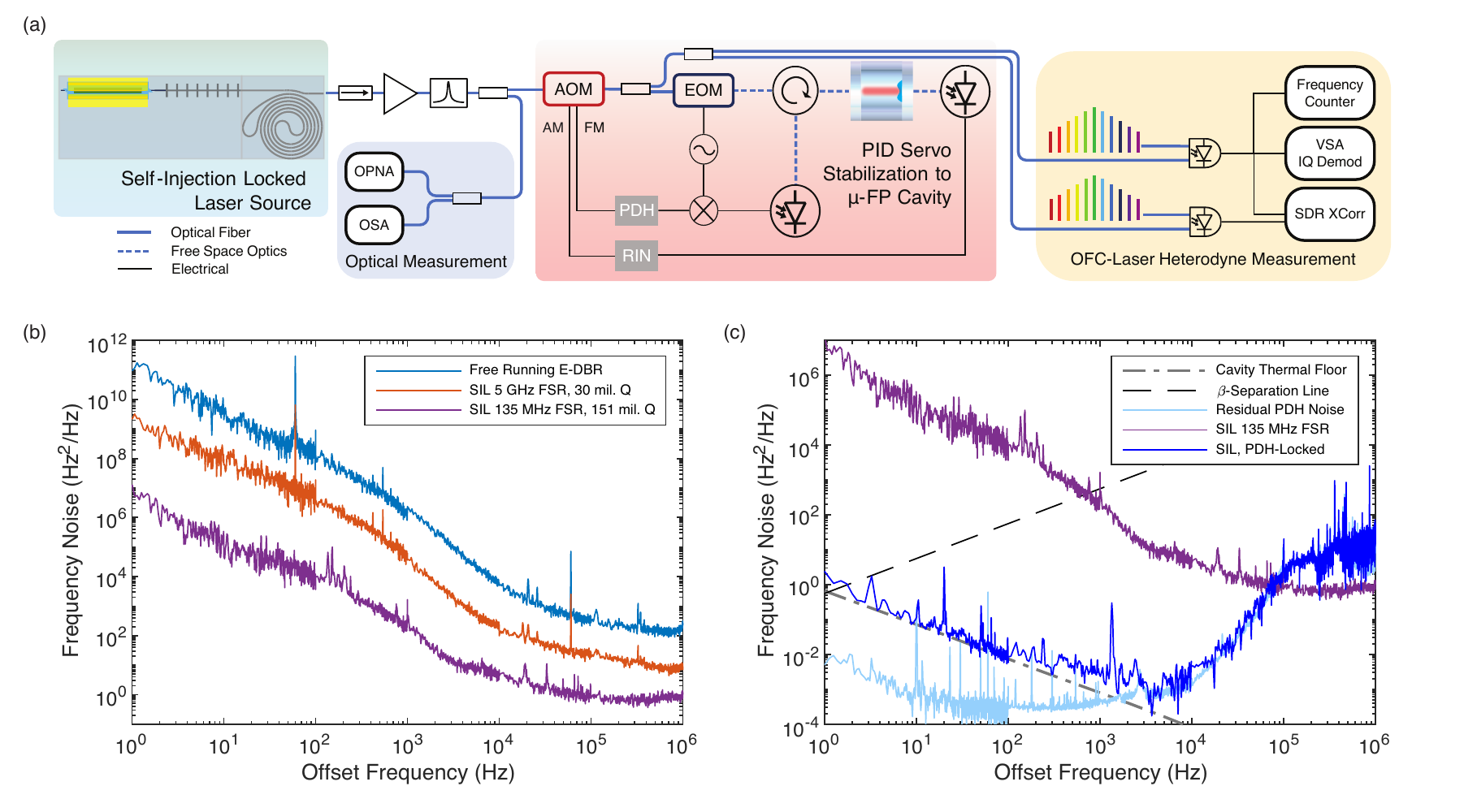}
\caption{Setup and frequency noise \textbf{(a)} The self-injection locked (SIL) laser output is split for optical spectrum analyzer (OSA) and frequency noise measurement with the optical phase noise analyzer (OPNA). Frequency and amplitude modulation (FM, AM) signals from the Pound-Drever-Hall (PDH) and relative intensity noise (RIN) lock servos, respectively, drive the acousto-optic modulator (AOM) to stabilize the SIL laser light. The resulting frequency noise, spectral lineshape, and Allan deviation are measured via heterodyne with a stabilized frequency comb. Two beat notes via separately stabilized combs are sampled for cross-correlation, providing greater measurement sensitivity above 1~kHz offset. \textbf{(b)} Frequency noise spectrum of the E-DBR laser free-running and self-injection locked, showing higher noise reduction by locking to a resonator with higher $Q$. \textbf{(c)} frequency noise spectrum of the best SIL and PDH-locked laser, revealing limits of the cavity thermal floor and residual PDH noise. The $\beta$-separation line shows the intersection with the frequency noise, used to estimate the integrated linewidth.}
\label{fig:FN}
\end{figure*}

\section*{Results}
To stabilize the laser frequency, the E-DBR laser wavelength is first tuned by current and temperature to match the nearest $\mu$-FP resonance. Figure \ref{fig:tuning} displays a tuning map of the E-DBR laser frequency with drive current and temperature. Once the laser and $\mu$-FP are aligned, the spiral resonator is also thermally tuned to self-injection lock the E-DBR laser at the same frequency and passively stabilize the laser. For further PDH-locking, the SIL laser is fiber coupled and stabilized to the $\mu$-FP with an acousto-optic modulator (AOM) as the frequency actuator. Figure \ref{fig:FN}a shows the laser stabilization and measurement setup. Further details are presented in the Materials and Methods section.

Frequency noise measurements are taken to compare self-injection locking to resonators with different $Q$s and to compare with servo locking (details in Materials and Methods). Figure \ref{fig:FN}b shows the frequency noise power spectral density (PSD) of the E-DBR laser free-running and self-injection locked to resonators with measured $Q$s and FSRs of 30~million at 5~GHz and 151~million at 135~MHz. Dependence of the TRN floor on modal volume (inversely proportional to FSR) of these resonators has been previously shown, as well as greater noise reduction with higher $Q$~\cite{Li:21, jin2021hertz}. 1/$f$ noise dominates at low offsets~\cite{van1988unified}. The SIL laser utilizing the 135~MHz FSR spiral resonator exhibits a hertz-level instantaneous linewidth and is used for subsequent PDH-locking. Its frequency noise is shown as the bottom trace in Figure \ref{fig:FN}b and the top trace in Figure \ref{fig:FN}c.  

Figure \ref{fig:FN}c shows the frequency noise of the SIL laser PDH-locked to the $\mu$-FP cavity. The measured noise follows the simulated cavity thermal floor closely out to kHz offsets, where the PDH-lock residual noise begins to take over (taken from the PDH servo in-loop error signal PSD). This diagnostic indicates that below kHz offsets, due to the stability of the SIL laser, the feedback loop can provide sufficient gain to lock to cavities with even lower noise floors. However, at offsets above 10 kHz, the feedback gain and phase begin to roll off. The phase roll-off is likely limited by the AOM and/or VCO, and a servo bump of 222~kHz is confirmed by turning up the servo proportional gain until oscillations are induced in the PDH signal. At even higher offset frequencies, the cross-spectrum data starts to become average limited. 

A distinct benefit of using our integrated SIL laser over a commercial fiber laser (besides the >10$\times$ smaller volume) is the lower frequency noise at high offsets. In atomic clock applications, the high offset noise can manifest via aliasing, which becomes increasingly critical at longer atom interrogation times~\cite{boyd2007high, ludlow2007compact}. In the case of PDH-locking to our $\mu$-FP with our SIL laser and with a commercial NKT fiber laser, there is up to a 10~dB difference in frequency noise at offset frequencies between 1~kHz to 1~MHz, despite the feedback loop roll-off~\cite{mclemore2022thermal}. Even lower high offset frequency noise can be obtained by using the drop port of the spiral resonator, which itself acts as a low-pass filter~\cite{Li:21, jin2021hertz}. Thus, these results demonstrate that our SIL laser can replace fiber lasers in high-finesse cavity locks, with even better performance at high offset frequencies. 

To preserve the hertz-level instantaneous linewidth of the SIL laser, greater feedback loop bandwidth is necessary. An EOM can be used as a frequency actuator with high derivative gain to enable frequency noise suppression out to multi-MHz offsets~\cite{endo2018residual}. Regardless, by PDH-locking the SIL laser, the frequency noise is reduced over 6 orders of magnitude at 1~Hz offset, 6 orders at 100~Hz, and 3 orders at 10~kHz. The corresponding phase noise at 100 Hz and 10 kHz offsets is about -65~dBc/Hz and -106~dBc/Hz, respectively, with a phase noise floor below -115~dBc/Hz. In a self-referenced optical frequency division (OFD) scheme, our laser can support 10~GHz microwave generation with phase noise of -150~dBc/Hz at 100~Hz offset; we note, however, the microwave phase noise floor is likely to be limited by photodetection rather than the laser phase noise floor~\cite{quinlan2013exploiting}.

\begin{figure}[htb]
\centering
\includegraphics[width=\linewidth]{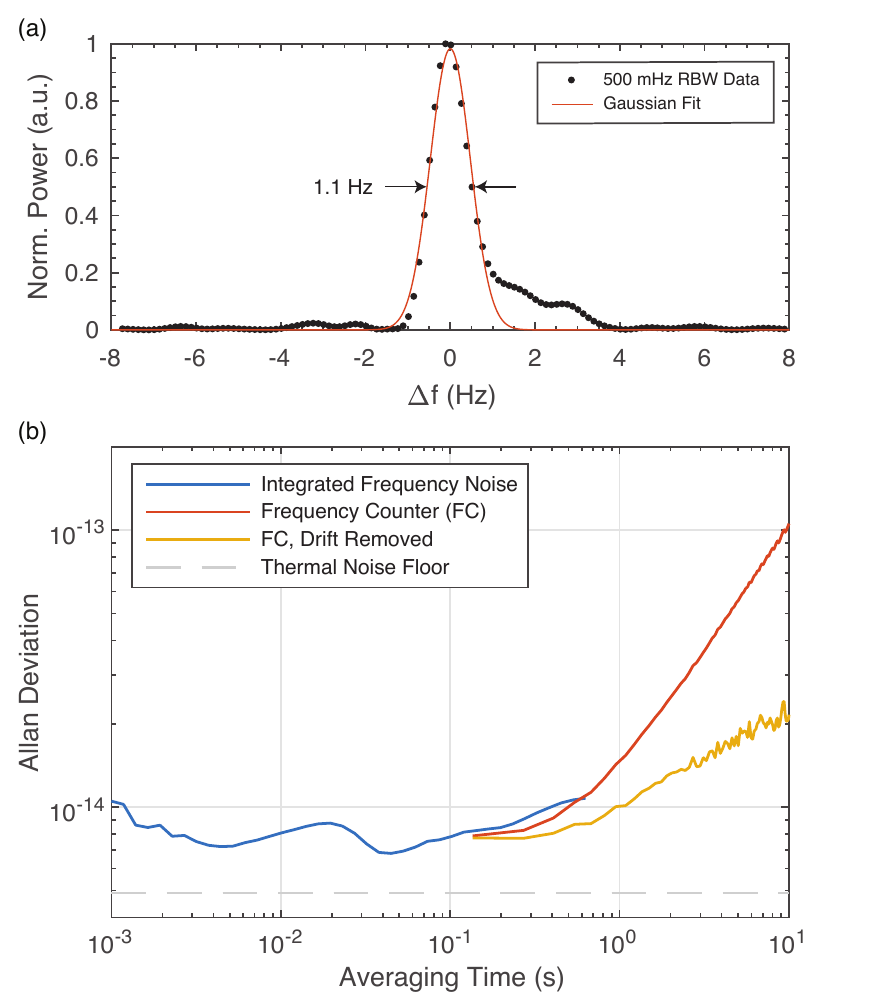}
\caption{RF spectrum and Allan deviation. \textbf{(a)} 500~mHz resolution bandwidth (RBW) RF spectrum of the laser heterodyne beat. A Gaussian fit reveals a 1.1~Hz FWHM, with 79\% of the power falling within the Gaussian. \textbf{(b)} Allan deviation taken by converting cross-spectrum frequency noise (integrated from 1~Hz to 10~kHz) for shorter averaging times and by frequency counter (0.1~s gate time) for longer averaging times. Frequency counter data with linear drift of 2.8~Hz/s removed reveals $10^{-14}$ fractional frequency stability out to 1~s.}
\label{fig:adev}
\end{figure}

In addition to frequency noise, the integrated laser linewidth is an important performance metric, particularly for spectroscopy of narrow atomic transitions. Estimating linewidth from frequency noise typically hinges on simplified assumptions of the PSD shape. We use two common definitions to estimate the integrated linewidth. The first estimation defines a $\beta$-separation line to separate high and low modulation index areas in the frequency noise PSD, which distinguishes contributions to the linewidth versus the wings of the lineshape~\cite{DiDomenico}. In our calculation, the frequency noise is integrated from 1~Hz out to the intersection with the $\beta$-separation line, yielding an integrated linewidth of 5.6~Hz on the timescale of 1~s. This likely represents an upper limit given the sparse sampling of the frequency noise spectrum near 1~Hz. Using this same method, the calculated integrated linewidths from the SIL laser and free-running E-DBR are 8.7~kHz and 1.4~MHz, respectively. 

Another method for estimating linewidth integrates the phase noise from high offset frequencies (short timescales) down to lower offset frequencies (up to longer timescales) until the integral exceeds $1/\pi$ rad$^2$~\cite{liang2015ultralow, hjelme1991semiconductor}. The corresponding offset frequency is defined as the linewidth. Cutoffs of 1 or $\pi^2$~rad$^2$ have also been previously defined, corresponding to an accumulated RMS phase deviation exceeding 1~radian or half a cycle, respectively~\cite{matei20171}. In our case, all give values on the same order. Using a cutoff of $1/\pi$ rad$^2$, we obtain linewidths of 8.9~kHz (free-running), 620~Hz (SIL), and 1.3~Hz (PDH). The disagreement between the $\beta$-separation line and phase integration methods in the free-running and SIL laser cases highlights the estimation oversimplification for nontrivial frequency noise PSD~\cite{brodnik2021fiber, liu202136}. However, in the case of our cavity-locked laser, these estimations agree on hertz-level linewidth. Yet another way to estimate linewidth is by measuring the width and shape of the power spectrum via heterodyne with a source having narrower linewidth. Shown in Figure \ref{fig:adev}a is an RF beat spectrum measured with 500~mHz resolution bandwidth on the VSA. Fitting the data and integrating the power reveals 79\% of the power falls within a Gaussian with a FWHM of 1.1~Hz. 

The Allan deviation is a measure of fractional frequency stability $\delta f/f$ as a function of averaging time. This is shown in Figure \ref{fig:adev}b, calculated for short averaging times by integrating the cross-spectrum frequency noise from 250~mHz to 10~kHz~\cite{liang2015ultralow}. For longer averaging times, frequency counter data (0.1 s gate time) is converted into Allan deviation with and without removal of a 2.8~Hz/s linear drift. Comparison of these two data sets around 200~ms averaging time shows good agreement, yielding a fractional frequency instability better than $10^{-14}$ from 1~ms to 1~s of averaging after drift removal. The result is an order of magnitude lower than what has been reported for the best integrated planar waveguide, fiber, and compact solid dielectric cavities~\cite{liu2020photonic, liu202136, loh2020operation, alnis2011thermal, zhang2020ultranarrow, stern2020ultra, liang2015ultralow} illustrating the clear advantage of our miniaturized vacuum-gap approach~\cite{mclemore2022thermal}. With a 1~s fractional frequency instability of 10$^{-14}$ and 1.1~Hz linewidth, our cavity-stabilized laser is capable of probing comparable 1~Hz atomic lines on the 1~s time scale and faster~\cite{ludlow2015optical, boyd2007high}.

\section*{Discussion}
In summary, we report a cavity-stabilized laser utilizing a planar semiconductor laser and lithographically formed micromirrors, with a 1.1~Hz linewidth on timescales of 1~s, 10$^{-3}$~Hz$^2$/Hz frequency noise floor at kHz offsets, and sub-10$^{-14}$ fractional frequency instability out to 1~s of averaging. Compared to the free-running laser, the PDH-locked SIL laser exhibits 11 orders of magnitude frequency noise reduction at 1~Hz offset. All critical components are lithographically manufactured devices with wafer-level scalability. The micro-fabricated mirrors used in this $\mu$-FP cavity highlight a path towards parallel fabrication of high-finesse reference cavities, and the small size of the micromirrors also allow for greater miniaturization of cavities in the future. Additionally, since our pump laser and spiral resonator share the same Si$_3$N$_4$ cross-sectional geometry, full integration of a heterogeneous self-injection locked laser is straightforward. Full integration not only reduces the laser-resonator coupling loss but also provides flexibility for on-chip semiconductor optical amplifier (SOA) integration to reach the output powers necessary for the aforementioned applications~\cite{davenport2016heterogeneous}. Adjusting the laser grating strength can also increase the output power~\cite{xiang2021high}. Lastly, integrated heaters can be deposited above the spiral resonator cladding, such that the SIL laser is made frequency agile~\cite{lihachev2021ultralow}.

While these results were obtained with the $\mu$-FP cavity separately housed, with our use of planar fabrication for key components, we foresee a path towards full integration. The AOM used in the PDH frequency lock may be eliminated in favor of thermal tuning of the SIL laser for slow, large dynamic range feedback and EOM actuation for high-speed control~\cite{endo2018residual}. The flexibility of the heterogeneous silicon photonics platform is well established, and future fully integrated III-V/Si/Si$_3$N$_4$ photonic circuits can include modulators and detectors to enable integrated PDH lock capabilities~\cite{xie2019heterogeneous, tran2017integrated, idjadi2017integrated, idjadi2020nanophotonic}. Passively held high vacuum in mm-scale laser cooled atomic systems has been demonstrated~\cite{boudot2020enhanced}, and we believe this can be adapted for vacuum-gap $\mu$-FP cavities. The laser and spiral resonator currently require a packaging volume of less than 50~mL, and we envision a customized heat shield and vacuum housing for the $\mu$-FP cavity of less than 100~mL; these are both over an order of magnitude smaller in volume compared to the previously discussed 900~mL commercial NKT fiber laser~\cite{mclemore2022thermal} and a 2.5~L aluminum vacuum chamber housing for a comparable 1~s 7.5 $\times$ 10$^{-15}$ Allan deviation 52~mL compact cavity~\cite{didier2018ultracompact}, respectively. As with all such ultrastable lasers, compact and robust vibration suppression is critical for operation outside a laboratory environment~\cite{thorpe2010measurement}, and will require further investigation. For further integration, coupling to the  $\mu$-FP cavity from a planar waveguide circuit via bonding could be achieved with metasurfaces and grating couplers, similar to those employed successfully in atomic systems~\cite{ropp2021multi, mcgehee2021magneto, spektor2021inverse}; alternatively, the cavity could be edge-coupled to the photonic circuit with a gradient index lens. Together with the remarkable progress in photonic integration, our work illustrates a path to a compact hybrid-integrated, cavity-stabilized laser package with 1~Hz integrated linewidth.

\newpage

\section*{Materials and Methods}
\noindent\textbf{Extended-DBR (E-DBR) Laser}

\noindent The heterogeneous III-V/Si/Si$_3$N$_4$ E-DBR laser was fabricated utilizing sequential wafer-bonding and processing of SOI and InP onto a 100~mm pre-processed Si$_3$N$_4$ wafer, with the lithographic alignment capabilities of a 248~nm deep-UV (DUV) stepper~\cite{xiang2021high}. 90~nm thick low-pressure chemical vapor deposition (LPCVD) Si$_3$N$_4$ was etched and cladded with deuterated SiO$_2$ to form low-loss Si$_3$N$_4$ waveguides and gratings~\cite{jin2020deuterated}. After planarization of the oxide cladding by chemical-mechanical polishing (CMP), a 60~mm~x~60~mm silicon-on-insulator (SOI) piece is then bonded, followed by substrate removal and Si processing to form the Si circuits. Cleaved InP-based MQW gain chiplets (grown on a 2-inch InP substrate) are then bonded, followed by substrate removal, III-V processing, oxide passivation, and metalization. 
The hybrid III-V/Si gain section is 1.5~mm long, and the 20~mm Si$_3$N$_4$ E-DBR is designed with a normalized grating strength $\kappa L_g$ of 1.75, resulting in a measured $\sim$5~GHz reflection bandwidth. The E-DBR feedback strength necessary for laser oscillation is enabled by low-loss intra-cavity III-V/Si and Si/Si$_3$N$_4$ mode converters as well as the low-loss Si$_3$N$_4$ grating~\cite{xiang2020narrow}. This laser was diced and packaged together with a thermistor and thermo-electric cooler (TEC) for edge-coupling to the resonator chip.

\smallskip

\noindent \textbf{Spiral Resonator}

\noindent The planar Si$_3$N$_4$ resonators are fabricated on 200~mm wafers at a CMOS foundry. The round-trip length of the spiral used for subsequent PDH-locking totals 1.41~meters, taking up an area of 9.2~mm~x~7.2~mm, limited by a 2~mm design minimum bend radius and 40~$\mu$m waveguide pitch for a 100~nm thick waveguide core~\cite{Li:21}. The loaded (intrinsic) $Q$ is measured to be 126~(164)~million, yielding an average propagation loss of 0.17~dB/m. These loss values are achieved by annealing at high temperatures over 1000~$^\circ C$ to drive out residual hydrogen~\cite{jin2021hertz}. After self-injection locking, the TRN-limited frequency noise is on par with that of commercial fiber lasers, which are typically necessary for PDH-locking to ultra-high-finesse cavities. In the experiment, the spiral resonator is placed on a thermally controlled stage to shift the resonance frequencies, and the phase is controlled via piezo-electric control of the gap between chips. 

\smallskip

\noindent \textbf{Micro-Fabricated Fabry-Perot ($\mu$-FP) Cavity}

\noindent The micromirror fabrication process consisted of patterning three disks of photoresist onto a super-polished glass substrate, which was then exposed to a solvent vapor reflow. As the solvent vapor was absorbed into the photoresist, the photoresist disks were reshaped, producing a small dimple on the top of each. The parabolic shape of this dimple served as a template for a concave mirror, which was then transferred to the substrate through reactive ion etch~\cite{jin2022scalable}. A single fused silica flat mirror forms the opposite mirror. After coating both mirror substrates with a highly reflective (>~99.999\% at 1550~nm) dielectric stack, they are optical-contact bonded to opposite sides of a 10~mm long, 25.4~mm diameter wide ULE glass spacer. With three concave micromirrors etched on one side, three distinct optical cavities were formed within a single test structure. For the locking experiment discussed here, only one optical mode is needed, so we use the micromirror with a 1.1~m radius of curvature, a finesse of 920~000 ($Q$ of 11.8~billion) and a linewidth of 16~kHz~\cite{mclemore2022thermal}. While the overall cavity volume is 8~mL, the cavity volume can be greatly reduced when constructed with only a single micromirror. Furthermore, the capability to fabricate multiple mirrors simultaneously on the same substrate could allow for the parallel manufacture of many single-micromirror cavities by bonding a substrate with an array of micromirrors, a spacer disk with a matching array of holes, and a flat mirror, then dicing the bonded stack. 

The length stability of the cavity is dominated by Brownian fluctuations in the dielectric mirror coatings at short timescales and thermal drift at longer timescales, which we minimize through cavity design and environmental isolation. Leveraging the versatility of the micromirror fabrication technique, we maximize the micromirror radius of curvature, leading to a large spot size on both end mirrors and effectively averaging stochastic cavity length fluctuations over a greater area. Fused silica for the mirror substrates contributes to a low noise floor through the material's high mechanical $Q$, while the ULE spacer reduces temperature sensitivity of the cavity mode. To further suppress thermal-induced drifts, the $\mu$-FP cavity is mounted in a custom heat shield inside a vacuum enclosure at $10^{-7}$~Torr, while temperature feedback is applied to the outside of the vacuum enclosure to ensure stability over long time periods. The result of these design and isolation considerations is a fundamental frequency noise floor of roughly $0.72/f$~Hz$^2$/Hz (phase noise floor of $-4.4/f^3$~dBc/Hz) and long-term cavity drift of a few Hz/s or better over hours-long time periods.

\smallskip

\noindent \textbf{Laser and Spiral Resonator Tuning}

\noindent As shown in Figure \ref{fig:tuning}, the laser frequency is tuned by varying the laser gain current and stage temperature. Single-mode and multi-mode states typical of DBR lasers are shown in each mode-hop tuning cycle, which were resolved with a high-resolution Apex optical spectrum analyzer (OSA)~\cite{xiang2021high, huang2019high}. For multi-mode states, only the strongest lasing mode is shown. To achieve stable locks, the laser is tuned to a single-mode state. We operated the laser at $\sim$232~mA and 22.5~$^\circ C$, due to the high single-mode output power at a frequency overlapping the nearest $\mu$-FP resonance. Sufficient tuning to cover over half the 15~GHz $\mu$-FP FSR via temperature control is shown, ensuring that given arbitrary frequency alignment of the laser and $\mu$-FP cavity, a $\mu$-FP resonance can always be reached. Once the laser frequency is tuned to the $\mu$-FP, the spiral resonator is cooled to $\sim$20.13~$^\circ C $, such that the resulting self-injection locked (SIL) laser frequency is centered on the $\mu$-FP cavity resonance. With an FSR of 135 MHz and tuning rate of $\sim$GHz/K, only modest thermal tuning is necessary~\cite{Li:21, jin2021hertz}. By placing the SIL laser setup in an enclosed box to shield from air currents, the SIL state is held for hours at a time.

\smallskip

\noindent \textbf{Stabilization Setup}

\noindent As Figure \ref{fig:FN}a depicts, the SIL laser is fiber coupled, isolated, amplified from 1~mW to 12~mW with a SOA, and filtered with a 1 nm optical band-pass filter. An acousto-optic modulator (AOM) serves as the frequency actuator in the feedback loop, and PDH sidebands are added with an electro-optic modulator (EOM). To interface the cavity, the EOM feeds to a free-space circulator and mode matching lens. After the AOM, the stabilized light is split off for measurement, leaving $\sim$500~$\mu$W incident on the cavity for locking. To reduce environmental noise, the cavity in its vacuum housing is mounted on an active vibration isolation platform inside a thermal and acoustic dampening enclosure. As in a typical PDH locking scheme, the sidebands mix with the carrier upon reflection from the cavity, such that the optical phase is photodetected~\cite{black2001introduction}. An error signal is retrieved after demodulating with a mixer, which is then filtered by a PID servo and fed back to a voltage-controlled oscillator to drive the AOM frequency. To further reduce frequency fluctuations below 1~kHz offsets, a relative intensity noise (RIN) servo is added to correct variations in the intra-cavity power. Optical intensity fluctuations in the cavity cause small shifts in the resonance frequency via local photo-thermal expansion in the mirrors, which have a small but finite absorption coefficient. Using the transmission detector to generate an error signal, the RIN servo applies amplitude modulation to the AOM RF power to stabilize the transmitted optical power.

\smallskip

\noindent \textbf{Frequency Noise Measurement}

\noindent Two separately verified methods for frequency noise measurement are utilized and shown in Figure~\ref{fig:FN}a. Frequency noise of the free-running and self-injection locked laser were taken using an OEwaves OE4000 optical phase noise analyzer (OPNA), based on a fiber Mach-Zehnder inteferometer (MZI) frequency discriminator. After PDH-locking to the $\mu$-FP cavity, a heterodyne beat is taken between the laser and a home-built optical frequency comb~\cite{nakamura2020coherent}, which is stabilized to a ytterbium lattice clock laser from an adjacent lab~\cite{mcgrew2018atomic}. The phase is extracted by IQ demodulation using an Agilent HP 89441A vector spectrum analyzer (VSA). The beat note is also used for frequency counter and spectral lineshape measurements. To enable higher measurement sensitivity at offset frequencies above 1~kHz, the stabilized light is split, sent along two physically separated fibers, and mixed with two independently stabilized frequency combs~\cite{nakamura2020coherent, kelleher2021centimeter}, so as to remove uncorrelated environmental and measurement noise, i.e., noise from the fibers or combs. The time-domain phase records of the two beat notes are simultaneously sampled at 2~MSa/s by software defined radio (SDR) and digitally cross-correlated to extract the common laser noise. In Fig.~\ref{fig:FN}c, the VSA and cross-spectrum SDR data are stitched together at 100~Hz offset frequency. 

\newpage

\bibliography{references}

\medskip
\begin{footnotesize}

\noindent \textbf{Funding}: 
This research was supported by DARPA GRYPHON, LUMOS, and APHI contracts HR0011-22-2-0009, HR0011-20-2-0044, and FA9453-19-C-0029 and the NIST on a Chip (NOAC) program.

\noindent \textbf{Acknowledgments}: 
The authors thank Andrew Ludlow's group (NIST-Boulder) for stable Ytterbium laser reference light, Paul Morton (Morton Photonics), Jonathan Peters (UCSB), and David Kinghorn (Pro Precision Process) for their respective contributions in the design, fabrication, and packaging of the E-DBR laser, and Martin Boyd (Vector Atomic) for discussions on optical clocks. Commercial equipment is identified for scientific clarity only and does not represent an endorsement by NIST.

\noindent \textbf{Data and materials availability}:  
All data needed to evaluate the conclusions in the paper are present in the paper. 
Additional data related to this paper may be requested from the authors. 

\end{footnotesize}

\end{document}